\documentclass{ws-procs9x6-cpt16}
\begin{document}

\newcommand{\refeq}[1]{(\ref{#1})}
\def\etal {{\it et al.}}

\title{Searches for Anisotropic Lorentz-Invariance Violation in the\\
 Photon Sector}

\author{F.\ Kislat and H.\ Krawczynski}

\address{Department of Physics and McDonnell Center for the Space Sciences\\
Washington University in St.~Louis, St.~Louis, MO 63130, USA}

\begin{abstract}
Lorentz invariance, the fundamental symmetry of Einstein's theory of Special Relativity, has been established and tested by many classical and modern experiments.
However, many theories that unify the Standard Model of particle physics and General Relativity predict a violation of Lorentz invariance at the Planck scale.
While this energy range cannot be reached by current experiments, minute deviations from Lorentz symmetry may be present at lower energies.
Astrophysical experiments are very suitable to search for these deviations, since their effects accumulate as photons travel across large distances.
In this paper, we describe astrophysical methods that we used to constrain the photon dispersion and vacuum birefringence.
\end{abstract}

\bodymatter

\section{Introduction}
The Standard-Model Extension (SME)\cite{sme} is an effective field theory approach to describe effects of a more fundamental theory beyond the Standard Model by introducing additional terms to the Standard-Model lagrangian.
In the photon sector, these additional terms result in the following dispersion relation:
\begin{equation}
  E(p) \simeq \left(1 - \varsigma^0 \pm \sqrt{\bigl(\varsigma^1\bigr)^2 + \bigl(\varsigma^2\bigr)^2 + \bigl(\varsigma^3\bigr)^2}\right) \, p,
\end{equation}
with
\begin{eqnarray}
  \label{eq:sigma0}\varsigma^0 &=& \sum_{djm}p^{d-4} Y_{jm}(\theta_k, \varphi_k)c_{(I)jm}^{(d)}, \\
  \varsigma^\pm &=& \varsigma^1 \pm \varsigma^2 = \sum_{djm}p^{d-4} {}_{\mp2}{Y}_{jm}(\theta_k, \varphi_k) \left(k_{(E)jm}^{(d)} \mp ik_{(B)jm}^{(d)}\right), \\
  \label{eq:sigma3}\varsigma^3 &=& \sum_{djm}p^{d-4} Y_{jm}(\theta_k, \varphi_k)k_{(V)jm}^{(d)}.
\end{eqnarray}
Hence, $\varsigma^0$ results in an energy and direction-dependent photon dispersion, and $\varsigma^\pm$ and $\varsigma^3$ additionally introduce a polarization dependence and thus vacuum birefringence.
Furthermore, Eqs.\ \refeq{eq:sigma0}-\refeq{eq:sigma3} imply a direction dependence of photon propagation.
The coefficients $c_{(I)jm}^{(d)}$, $k_{(E)jm}^{(d)}$, and $k_{(B)jm}^{(d)}$ are nonzero only for even $d$, while $k_{(V)jm}^{(d)}$ are nonzero only for odd $d$, where $d$ is the mass dimension of the corresponding operator.

Astrophysical tests are highly sensitive even to tiny values of these coefficients since their effects accumulate as photons travel across large distances.
Photon dispersion is tested by measuring arrival times of photons from short transient events such as gamma-ray bursts at different wavelengths; see, e.g., Ref.\ \refcite{fermi-grbs}.
Vacuum birefringence induced by $\varsigma^\pm$ and $\varsigma^3$ leads to an energy-dependent rotation of the polarization vector, which can be observed directly in spectropolarimetric measurements.

Due to the anisotropic nature of Lorentz-invariance violation (LIV) in the SME, searches utilizing a single source can only test a linear combination of coefficients.
For example, at $d=5$ there are 16 real coefficients determining the complex $k_{(V)jm}^{(5)}$, and at $d=6$ there are 25 real coefficients of the $c_{(I)jm}^{(6)}$.
In this paper we present a search for anisotropic LIV using optical polarization measurements of active galactic nuclei (AGNs),\cite{kislat_f_krawczynski_h_2016} and a search for anisotropic nonbirefringent LIV using gamma-ray time-of-flight measurements of AGN flares.\cite{kislat_f_krawczynski_h_2015}
In both searches, we observe multiple astrophysical sources and then constrain the coefficients $k_{(V)jm}^{(5)}$ and $c_{(I)jm}^{(6)}$ individually using a spherical decomposition of the results.
Here we outline the methods used, while the actual limits on the SME coefficients will be published elsewhere.
X-ray polarization measurements of gamma-ray bursts have already been used to constrain the $d=5$ coefficients in previous studies.\cite{va_kostelecky_m_mewes_2013,xray-polarimetry}
However, the statistical and sytematic errors on the reported X-ray polarization properties are very large and the detections are still not firmly established.

\section{Optical polarimetry}\label{sec:polarimetry}
The polarization angles of two photons observed at energies $E_1$ and $E_2$ emitted at redshift $z_k$ which initially have the same polarization angle will differ by\cite{va_kostelecky_m_mewes_2013}
\begin{equation}\label{eq:delta_psi}
    \Delta\psi = 
        (E_1^{\,2} - E_2^{\,2}) \, L_{z_k}^{(5)} \sum_{\substack{j = 0 \ldots 3    \\
                                                                 m = -j \ldots j}}
        Y_{jm}(\theta_k,\varphi_k)k_{(V)jm}^{(5)} \\
    \equiv (E_1^{\,2} - E_2^{\,2}) \, \zeta_k^{(5)},
\end{equation}
where we introduced the parameter~$\zeta_k^{(5)}$, assuming that $d=5$ terms are the dominant correction to the Standard-Model lagrangian.
We measured $\Delta\psi$ using multiple spectropolarimetric observations of 27 different AGNs in the northern hemisphere.
Using a likelihood ratio test, we compared the observations of distant ($z > 0.6$) to ``nearby'' ($z < 0.4$) sources in order to set upper limits on the LIV-induced rotation parameter~$\zeta_k^{(5)}$.

Furthermore, we used spectrally integrated optical polarization measurements of 36 southern-hemisphere AGNs.
The rotation of the polarization angle leads to a partial cancellation of the net polarization.
For each observation, we found the largest value of~$\zeta_k^{(5)}$ that was still in agreement with the observed polarization fraction, as an upper limit on a possible LIV.

Both the spectropolarimetric and the spectrally integrated polarization measurements resulted in strong upper limits on LIV in the photon sector for each source.

\section{Gamma-ray time-of-flight measurements}\label{sec:timeofflight}
While most coefficients of the SME are best constrained using polarization measurements, the coefficients $c_{(I)jm}^{(d)}$ with even $d$ are nonbirefringent.
They can only be constrained using time-of-flight measurements.
We used measurements of $500\,\mathrm{MeV}$ to $300\,\mathrm{GeV}$ gamma-ray lightcurves of 24 AGNs to derive direction-dependent constraints on the speed of light in vacuum.\cite{kislat_f_krawczynski_h_2015}
Using the \emph{DisCan} method,\cite{discan} we found constraints on the quadratic energy dependence of the speed of light introduced by the $d=6$ SME coefficients.

No significant energy dependence of the light-travel time was found, and we set upper limits $\gamma_k^{(6)}$ on the LIV coefficients.
They are not strong enough to constrain LIV at the Planck scale.
However, they represent the first complete set of constraints in the $d=6$ photon sector.

\section{Anisotropic Lorentz-invariance violation}\label{sec:anisotropic}
When constraining the rotation of the polarization direction or the energy-dependence of the photon velocity from an astrophysical source $k$, a limit~$\gamma_k$ is placed on a linear combination of SME coefficients:
\begin{equation}
  \Bigl|\sum_{\substack{j = 0 \ldots 3 \\ m = -j \ldots j}} Y_{jm}(\theta_k,\varphi_k)k_{(V)jm}^{(5)}\Bigr| < \gamma_{k}^{(5)}
  \;\text{or}\;
  \Bigl|\sum_{\substack{j = 0 \ldots 4 \\ m = -j \ldots j}} Y_{jm}(\theta_k,\varphi_k)c_{(I)jm}^{(6)}\Bigr| < \gamma_{k}^{(6)}.
\end{equation}
When observing $N$ sources, we can rewrite these inequalities in matrix form in terms of the real parameters comprising the SME coefficients:
\begin{equation}\label{eq:gamma_linear_combination}
  \mathbf{H} \bullet \boldsymbol{v} < \boldsymbol{\gamma},
\end{equation}
where $\boldsymbol{v}$ is a vector of the components of the $k_{(V)jm}^{(5)}$ or $c_{(I)jm}^{(6)}$, $\boldsymbol{\gamma}$ is a vector of the values $\gamma_k^{(d)}$, and $\mathbf{H}$ is the $M \times N$ coefficient matrix (with $M = 16$ for $d=5$, and $M=25$ in the $d=6$ case).
We find limits on the components of $\boldsymbol{v}$ using Monte Carlo integration: we sampled $10^7$ random vectors $\boldsymbol{\gamma}$ by drawing each components from a normal distribution with a standard deviation chosen such that the values satisfy the confidence level of the limits $\gamma_k^{(d)}$.
We then solve the equality corresponding to Eq.~\refeq{eq:gamma_linear_combination}:
\begin{equation}
  \boldsymbol{v} = (\mathbf{H}^T\mathbf{H})^{-1}\mathbf{H}^T\boldsymbol{\gamma}.
\end{equation}
Each solution marks a point in the space of the SME coefficients.
In this way, we build the distribution of SME coefficients and find their $95\%$ upper and lower bounds.
The actual coefficient constraints will be published in Refs.\ \refcite{kislat_f_krawczynski_h_2016} and\ \refcite{kislat_f_krawczynski_h_2015}.

\section{Summary}
In order to constrain the $c_{(I)jm}^{(6)}$ at the Planck scale, time-of-flight measurements of photons with energies of the order of $100\,\mathrm{PeV}$ would be necessary,\cite{kislat_f_krawczynski_h_2015} which is currently not possible.
On the other hand, it may be possible with future gamma-ray instruments to measure gamma-ray polarization at~$100\,\mathrm{MeV}$, which would allow us to constrain the coefficients $k_{(E)jm}^{(6)}$ and $k_{(B)jm}^{(6)}$.
The next step, however, will be to constrain the birefringent coefficients of mass dimension $d=4$, which currently are not fully constrained.\cite{datatables}

\end{document}